# Bifurcations in Globally Coupled Map Lattices

Wolfram Just
Theoretische Festkörperphysik
Technische Hochschule Darmstadt
Hochschulstraße 8
D–64289 Darmstadt
Germany

April 28, 1994

**Abstract**

The dynamics of globally coupled map lattices can be described in terms of a nonlinear Frobenius–Perron equation in the limit of large system size. This approach allows for an analytical computation of stationary states and their stability. The complete bifurcation behaviour of coupled tent maps near the chaotic band merging point is presented. Furthermore the time independent states of coupled logistic equations are analyzed. The bifurcation diagram of the uncoupled map carries over to the map lattice. The analytical results are supplemented with numerical simulations.

# 1  Introduction

A lot of physical non equilibrium systems are governed by the motion of many relevant degrees of freedom. The most prominent example is hydrodynamic turbulence but also optical, magnetic, chemical and biological systems fall into this class (e.g [1] and references therein). From the theoretical point of view it is often too difficult to solve the equations of motion directly even on the largest accessible computers. Hence one is forced to investigate mathematical model systems especially if one is interested in the pattern formation out of a random state. One class of such systems is given by coupled simple maps [2]. Several phenomena arising in the context of hydrodynamics, optics and solid state physics have been treated by such models [3, 4]. Although the actual relation between the basic equations of motion and the map lattice model is often not very well justified the simplicity of these models allow for a detailed numerical investigation. Furthermore rigorous approaches can be applied [5, 6, 7]. Hence it may be possible to understand the basic mechanisms of pattern formation from a random state. But the general theory is just at the beginning. For that reason it is desirable to investigate the dynamics of simple spatially extended model systems beyond numerical simulations.

Frequently coupled map lattices with short range interaction are investigated. This kind of interaction models the diffusive character of the spatially extended system. But it seems to be too difficult to investigate theoretically. A much more simpler type is given by an all to all interaction. This coupling may be considered as a limiting case of a long range interaction [8] or as an approximation of a short ranged coupled system in a (spatially) high dimensional lattice [9]. Furthermore such type of coupling has been used to model e.g. the dynamics in multimode lasers [10], of Josephson junction arrays [11], or the dynamics in certain biological systems [12]. Its mean field like character greatly simplifies the theoretical approach [13, 14]. Hence I arrive at the model equation

$$x_{n+1}^{(\nu)} = (1-\epsilon)f\left(x_n^{(\nu)}\right) + \frac{\epsilon}{N}\sum_{\mu=0}^{N-1} f\left(x_n^{(\mu)}\right) \quad . \tag{1}$$

Here $f$ denotes the single site map which should admit a chaotic motion, $\epsilon$ the coupling constant, and $N$ the systems size. The coupling is brought about by the mean field

$$h_n = \frac{1}{N}\sum_{\mu=0}^{N-1} f\left(x_n^{(\mu)}\right) \quad . \tag{2}$$

Although much more general types of globally coupled systems may be considered I restrict the discussion to eq.(1). The limit of large system size $N \gg 1$ is of special interest because in that case the number of degrees of freedom which are relevant for the time evolution may become large. Some surprising feature of global quantities like the mean field (2) has been discovered recently [15]. The mean square deviation does not decrease with the system size but saturates at a finite value. This phenomenon has been termed "violation of the law of large numbers" and has been attributed to some hidden coherence in the coupled map lattice [16, 17, 18, 19, 20, 7].

The limit of large system size causes some problems. On one hand a direct numerical simulation of eq.(1) needs increasing computer capacities. On the other hand the occurrence of long transients considerably complicates the study of stationary states [21, 7]. For that reason an alternative approach is desirable. Because of the global coupling the systems allows for a description in terms of a reduced density [16, 22, 7]. Consider the probability that a phase space coordinate takes the value $x$. It reads

$$\rho_n(x) = \frac{1}{N} \sum_{\nu=0}^{N-1} \delta\left(x - x_n^{(\nu)}\right) \quad . \tag{3}$$

The mean field can be expressed in terms of this density via

$$h_n = \int f(x)\rho_n(x)\,dx \quad . \tag{4}$$

For the time evolution of the density (3) one obtains from eq.(1)

$$\rho_{n+1}(x) = \int \delta\left(x - T_n(y)\right)\rho_n(y)\,dy \tag{5}$$

where the mean field map $T_n$ is given by

$$T_n = F_n \circ f \tag{6}$$
$$F_n(x) = (1-\epsilon)x + \epsilon h_n \quad . \tag{7}$$

Eqs.(4), (5), (6) and (7) constitute a closed set of exact nonlinear evolution equations which is easier to handle as the original coupled map lattice. The system size enters in this formulation only via the structure of the density (3). In the limit of infinite system size the density tends for typical phase space points to a "sufficiently continuous" function[1] (cf. ref. [23] for related rigorous statements). Therefore the limit of large system size is contained in this mean field description in a quite simple way. I restrict the analysis in the sequel to the discussion of eq.(5) and continuous densities.

Let me focus on the investigation of stationary states, that means fixed points and periodic orbits of the mean field equation and their stability. For simplicity in the notation the explicit formulation will be given for fixed points only. The general considerations are easily extended to periodic orbits by investigating the fixed point problem of a suitable iterate. The fixed points are determined by

$$\rho_*(x) = \int \delta\left(x - T_*(y)\right) \rho_*(y) \, dy \tag{8}$$

where $T_*$ denotes the mean field map evaluated at the fixed point $\rho_*$. Without specifying the single site map less can be said about the structure of the solution. However some conclusions can be drawn about the stability. Consider a small deviation from the fixed point density $\rho_n(x) = \rho_*(x) + \delta\rho_n(x)$. As long as linear stability analysis can be applied these deviations obey (cf. eq.(5))

$$\delta\rho_{n+1}(x) = \int \delta\left(x - T_*(y)\right) \delta\rho_n(y) \, dy - \delta T_n \frac{d}{dx}\rho_*(x) \tag{9}$$

$$\delta T_n = \epsilon \int f(x)\delta\rho_n(x) \, dx \quad . \tag{10}$$

---

[1] By this I mean that the corresponding measure $\mu_n(g) := \int g(x)\rho_n(x)\,dx$, $\forall g$, tends to a limit which has a non atomic component. Although the formulation in terms of measures is physically less appealing one should keep in mind that equations like eq.(5) have to be understood in the weak sense, $\mu_{n+1}(g) = \mu_n(g \circ T_n)$, $\forall g$, on a rigorous level. Especially the formal derivative $\rho'_n(x)$ has to be identified with the measure $\nu_n(g) := -\int g'(x)\rho_n(x)\,dx$, $\forall g$, even if the derivative of the density is not defined.

Stability is determined by the eigenvalue problem which corresponds to this linear evolution equation. Obviously this eigenvalue problem has the form of a perturbed Frobenius–Perron equation for the map $T_*$. The second term in eq.(9) yields the formal perturbation. It is well known that the Frobenius–Perron equation has an eigenvalue one which is the largest in modulus. Because of a symmetry of the full evolution equation (5) and the normalization of the density this eigenvalue is persistent with respect to the perturbation [7]. The corresponding eigenfunction may be understood as a kind of Goldstone mode. If the map $T_*$ has certain hyperbolic properties then the largest eigenvalue is isolated. Hence one needs a finite coupling strength in order that additional eigenvalues cross the unit circle. It is therefore expected that stationary states are stable in hyperbolic systems at least for small coupling strength. On the contrary the spectrum of the Frobenius–Perron operator for non hyperbolic maps is usually degenerated on the unit circle [24, 25]. Therefore an infinitesimal coupling may induce instability and a complicated dynamical behaviour.

On this level the considerations on stationary states and their stability is only qualitative. In section 2 the stability analysis will be put on a more rigorous basis. The explicit computation of stationary states requires the knowledge of the single site map $f$. In section 3 the analysis will be presented for the "hyperbolic" tent map which includes the bifurcations that occur in this simple system. The problem of non hyperbolic systems is analyzed in the context of the logistic equation. A partial bifurcation analysis and numerical simulations are presented in section 4. Finally I discuss the implications for the dynamics of the original coupled map lattice .

## 2  Stability of stationary solutions

The stationary solutions depend on the special map lattice under consideration. But the stability can be discussed quite generally if the existence of the fixed point is presupposed. Therefore the computation of the stationary states is postponed to the subsequent sections. Here considerations concerning their stability are presented.

I suppose that eq.(8) has been solved for a continuous density $\rho_*$ which should coincide with the Sinai–Ruelle–Bowen (SRB) measure of the map $T_*$. The instability of such a distribution is indicated by an exponentially increasing solution of eqs.(9) and (10). Its formal solution reads[2]

$$\delta\rho_n(x) = \int \delta\left(x - T_*^n(y)\right) \delta\rho_0(y)\, dy$$
$$- \sum_{\sigma=0}^{n-1} \delta T_{n-1-\sigma} \frac{d}{dx} \int \delta\left(x - T_*^\sigma(y)\right) (T_*^\sigma)'(y) \rho_*(y)\, dy \quad (11)$$

$$\delta T_n = \epsilon \int f\left(T_*^n(y)\right) \rho_*(y)\, dy$$
$$+ \epsilon \sum_{\sigma=0}^{n-1} \delta T_{n-1-\sigma} \int \frac{d}{dx} f\left(T_*^\sigma(x)\right) \rho_*(x)\, dx \quad (12)$$

For any continuous initial condition $\delta\rho_0$ the first term of eqs.(11) and (12) decays to zero because the SRB measure attracts any continuous distribution. Therefore an exponential increase, that means instability, can be excluded if for all sufficiently smooth functions $g$ and $h$ the quantity

$$J_n := \int \frac{d}{dx} g\left(T_*^n(x)\right) \cdot h(x) \rho_*(x)\, dx \quad (13)$$

remains bounded if $n$ tends to infinity. The subsequent considerations focus on the discussion of this quantity.

The expression (13) has a close relation to the linear response of the map $T_*$ [26, 27]. To clarify this point consider for the moment the map $T_*$ and denote its SRB density by $\rho_*$. Let $x_0 = x + \epsilon h(x)$ be an ensemble of initial points where the values $x$ are distributed according to the SRB measure. The time evolution of the expectation value of a smooth observable $g$ reads

$$\langle g \rangle_n = \langle g\left(T_*^n[x + \epsilon h(x)]\right) \rangle_* \quad (14)$$

where the brackets $\langle \ldots \rangle_*$ denote the SRB average. The quantity (14) tends towards the stationary value $\langle g \rangle_*$ in the limit $n \to \infty$. Expansion in the small quantity $\epsilon$ leads to

$$\langle g \rangle_n - \langle g \rangle_* = \epsilon \langle \frac{d}{dx} g\left(T_*^n(x)\right) \cdot h(x) \rangle_* + O(\epsilon^2) \quad . \quad (15)$$

---

[2] In the weak sense the relation (11) reads $\delta\mu_n(g) = \delta\mu_0(g \circ T_*^n) - \sum_{\sigma=0}^{n-1} \delta T_{n-1-\sigma} \mu_*((g \circ T_*^\sigma)')$ .

The right hand side yields the formal linear relaxation function which coincides with the expression (13). As long as the expansion in $\epsilon$ is uniformly valid in $n$ this quantity decays because the left hand side does. But a linear response theory is in general not valid in the context of low dimensional maps even for hyperbolic systems. Nevertheless any system which allows for the application of linear response theory leads to a bounded relaxation function (13) and therefore to stable stationary solutions in globally coupled maps. But this condition is too strong for general considerations.

The estimation of the quantity (13) requires the knowledge of the density $\rho_*$. It can be computed for the class of piecewise linear Markov maps. This kind of systems is characterized by maps of the interval $I$ which are linear on a suitable partition $I_\nu = [a_\nu, a_{\nu+1}]$, $0 \leq \nu \leq M$, $I = \cup_\nu I_\nu$, $\gamma_\nu = T'_*(x)$ for $x \in I_\nu$. Furthermore it is required that the image of each interval is a union of other intervals $T_*(I_\nu) = \cup'_\mu I_\mu$. This property implies the expansiveness of the map. Although these systems seem to look rather special a lot of one dimensional maps can be described in terms of such models by considering fine partitions $I_\nu$ of the phase space $I$ [28, 25]. It is well known that the dynamics of such systems is equivalent to a subshift of finite type and that the invariant density is a piecewise constant function, $\rho_*(x) = \rho_\nu$ for $x \in I_\nu$. Integration by parts immediately yields for the relaxation function (13)

$$J_n = \int g\left(T_*^n(x)\right) h'(x) \rho_*(x)\, dx + \sum_{\nu=1}^{M-1} g\left(T_*^n(a_\nu)\right) h(a_\nu) (\rho_\nu - \rho_{\nu-1}) \qquad (16)$$

The relaxation function is obviously a bounded function. Whereas the first contribution decays to a constant the second term is periodic for large $n$.

The considerations of this section imply that the stationary states of the mean field equation are stable if the mean field map $T_*$ can be cast into the form of a piecewise linear Markov map. At this level nothing can be said about non hyperbolic systems like the frequently discussed logistic equation. In this case two different situations have to be considered. If the map $T_*$ possesses a stable periodic orbit then the corresponding SRB density consists of a finite number of $\delta$ peaks. But then the factors $(T_*^\sigma)'(x)$ in eqs.(11) and (12) decay exponentially and imply the stability of the solution at least if the initial disturbance $\delta\rho_0$ is contained in the basin of attraction of the periodic orbit. This result is obviously trivial. In the other case the SRB density develops square root singularities[3]. But these singularities induce an exponential increase of the relaxation function (13) [26] and turn the stationary solution unstable. This fact corresponds to the violation of linear response theory because of the structural unstable character of the map. Hence one might expect that a small coupling $\epsilon$ induces a non stationary behaviour in the map lattice. We will come back to this phenomenon in section 4.

---

[3] This statement is rigorous if the orbit of the critical point has finite length.

# 3 Stationary states for coupled tent maps

A very simple but nontrivial model system is given by coupled tent maps

$$f(x) = 1 - a|x| \quad . \tag{17}$$

The isolated map has a chaotic motion for $1 < a \leq 2$ and shows a cascade of chaotic band merging at parameter values $a_n = 2^{1/2^n}$. The mean field map (6) which governs the dynamics of the globally coupled map lattice reads

$$T_n(x) = 1 - \epsilon a \langle |x| \rangle_n - (1-\epsilon)a|x| \tag{18}$$

where the brackets $\langle \ldots \rangle_n$ denote the average with respect to the density $\rho_n(x)$.

For a dense set of parameter values $a, \epsilon$ one can manage that the orbit of the extremal point $x = 0$ terminates at some unstable periodic orbit. The corresponding finite set of orbit points yields a Markov partition. The results of section 2 imply that any smooth periodic solution of the mean field equation is stable. Hence only the construction of these solutions from eq.(8) is desired.

For that purpose it is useful to make a linear time dependent scale transformation $x = \gamma_n z$ in the mean field equation (5) in order that the mean field map (18) takes the form of the uncoupled map (17). The evolution equation for the scaled density

$$\tilde{\rho}_n(z) = |\gamma_n|\rho_n(\gamma_n z) \tag{19}$$

reads

$$\tilde{\rho}_{n+1}(z) = \int \delta\left(z - \tilde{T}_n(z')\right) \tilde{\rho}(z') \, dz' \tag{20}$$

where the scaled mean field map is given by

$$\tilde{T}_n(z) = 1 - \frac{(1-\epsilon)a|\gamma_n|}{1 - \epsilon a|\gamma_n|\langle |z| \rangle_n}|z| \tag{21}$$

if the scaling constants are determined by

$$\gamma_{n+1} = 1 - \epsilon a|\gamma_n|\langle |z| \rangle_n \quad . \tag{22}$$

In the sequel the brackets denote the average with respect to the scaled density (19).

*Fixed points*: For time independent solutions $\tilde{\rho}_n = \tilde{\rho}_*$ eqs.(21) and (22) yield

$$\gamma_* = 1/(1 + \epsilon a \langle |z| \rangle_*) > 0 \tag{23}$$
$$\tilde{T}_*(z) = 1 - (1-\epsilon)a|z| \quad . \tag{24}$$

The actual shape of the fixed point density $\tilde{\rho}_*$ is determined by an equation analogous to eq.(8). It coincides with the smooth invariant density of the tent map at a parameter value $a_{red.} := (1-\epsilon)a$. The global coupling has the effect to reduce the parameter in the tent map.

*Period 2 solution*: The single tent map (17) admits a period two chaotic solution for $\sqrt[4]{2} < a < \sqrt{2}$. A corresponding period two orbit $\tilde{\rho}_1^*, \tilde{\rho}_2^*$ may be expected in the mean field dynamics (20) too. It is determined by the following set of coupled equations

$$\tilde{\rho}_1^*(z) = \int \delta\left(z - \tilde{T}_2^*(z')\right) \tilde{\rho}_2^*(z') \, dz'$$
$$= \int \delta\left(z - (\tilde{T}_2^* \circ \tilde{T}_1^*)(z')\right) \tilde{\rho}_1^*(z') \, dz' \quad (25)$$
$$\tilde{T}_1^*(z) = 1 - \frac{a_{red.}|\gamma_1^*|}{\gamma_2^*}|z|, \quad 1 \leftrightarrow 2 \quad (26)$$
$$\gamma_2^* = 1 - \epsilon a |\gamma_1^*| \langle |z| \rangle_1^*, \quad 1 \leftrightarrow 2 \quad (27)$$

where $\langle \ldots \rangle_{1/2}^*$ denote the average with respect to the densities $\tilde{\rho}_{1/2}^*$.

First of all let me investigate the form of the general solution of eq.(25). For that purpose the ratio $c := \gamma_2^*/\gamma_1^*$ is considered as a free parameter which will be fixed later. The two times iterated mean field map which determines the shape of the density is given by

$$(\tilde{T}_2^* \circ \tilde{T}_1^*)(z) = 1 - a_{red.}|c - a_{red.}|z|| \quad . \quad (28)$$

Here $\gamma_i^* > 0$ has been assumed which is justified a posteriori. For $a_{red.} > \sqrt{2}$ the map admits only one continuous invariant density. It corresponds to the fixed point solution discovered above. Therefore let me concentrate on the opposite case. For $\sqrt[4]{2} < a_{red.} < \sqrt{2}$ and $c \in [1/a_{red.}, a_{red.}]$ the map decomposes into two different ergodic components which are separated by the unstable fixed point (cf. Fig.1)   <Fig.1

$$\zeta(c) = \frac{a_{red.}c - 1}{(a_{red.})^2 - 1} \quad . \quad (29)$$

The corresponding ergodic invariant densities are denoted by $\rho^{(-)}(z;c), |z| < \zeta(c)$ respectively $\rho^{(+)}(z;c), z > \zeta(c)$. It is an elementary but important geometric property of the map (28) that the variation of the parameter $c$ shifts the fixed point $\zeta(c)$ but leaves the two components of the map unchanged up to a linear scale transformation. Hence the dependence of the densities on the parameter $c$ can be written as

$$\rho^{(-)}(z;c) = \frac{\zeta(c')}{\zeta(c)} \rho^{(-)}\left(\frac{\zeta(c')}{\zeta(c)} z; c'\right), \quad |z| < \zeta(c) \quad (30)$$
$$\rho^{(+)}(z;c) = \frac{1 - \zeta(c')}{1 - \zeta(c)} \rho^{(+)}\left(1 - \frac{1 - \zeta(c')}{1 - \zeta(c)}(1-z); c'\right), \quad z > \zeta(c) \quad . \quad (31)$$

In addition both densities are not independent of each other. The geometric properties of the map (26) imply that (appendix A)

$$\int \delta\left(x - \tilde{T}_1^*(y)\right) \rho^{(\pm)}(y;c)\, dy = \rho^{(\mp)}\left(x; \frac{1}{c}\right) \tag{32}$$

holds. The general solution of eq.(25) is an arbitrary convex combination of both ergodic components

$$\tilde{\rho}_1^*(z) = (1-\alpha)\rho^{(-)}(z;c) + \alpha\rho^{(+)}(z;c), \quad \alpha \in [0,1] \quad . \tag{33}$$

But the parameter $c$ is not at our disposal. It has to satisfy the additional constraints (27). Because one has control of the $c$ dependence of the solution via eqs.(30) and (31) it will be shown that the relative weight $\alpha$ of the two ergodic components determines this ratio. To this end let me first mention that thanks to eqs.(32) and (33) the relations

$$\langle|z|\rangle_1^* = (1-\alpha)\langle|z|\rangle^{(-)}(c) + \alpha\langle|z|\rangle^{(+)}(c) \tag{34}$$

$$\langle|z|\rangle_2^* = (1-\alpha)\langle|z|\rangle^{(+)}(1/c) + \alpha\langle|z|\rangle^{(-)}(1/c) \tag{35}$$

hold where the abbreviations

$$\langle|z|\rangle^{(\pm)}(c) := \int |z| \rho^{(\pm)}(z;c)\, dz \tag{36}$$

have been used. Then eqs.(27) read

$$1/\gamma_1^* = c + \epsilon a\left[(1-\alpha)\langle|z|\rangle^{(-)}(c) + \alpha\langle|z|\rangle^{(+)}(c)\right] \tag{37}$$

$$1/\gamma_2^* = 1/c + \epsilon a\left[(1-\alpha)\langle|z|\rangle^{(+)}(1/c) + \alpha\langle|z|\rangle^{(-)}(1/c)\right] \tag{38}$$

which result in

$$1 - c = \epsilon a\left[(1-\alpha)\left(\langle|z|\rangle^{(-)}(c) - c\langle|z|\rangle^{(+)}(1/c)\right) + \alpha\left(\langle|z|\rangle^{(+)}(c) - c\langle|z|\rangle^{(-)}(1/c)\right)\right] \quad . \tag{39}$$

Because of the scaling property (30) and (31) the dependence of the expectation values on the parameter $c$ can be evaluated to

$$\langle|z|\rangle^{(-)}(c) = \frac{\zeta(c)}{\zeta(1)}\langle|z|\rangle^{(-)}(1) \tag{40}$$

$$\langle|z|\rangle^{(+)}(c) = 1 - \frac{1-\zeta(c)}{1-\zeta(1)}\left(1 - \langle|z|\rangle^{(+)}(1)\right) \quad . \tag{41}$$

Here $c'$ has been set to unity for convenience. If one additionally uses the relation $\langle|z|\rangle^{(+)}(1) = 1 - a_{red.}\langle|z|\rangle^{(-)}(1)$ between both expectation values (cf. appendix A) one ends up with

$$c = 1 + \frac{\epsilon a(a_{red.} - 1)(1 - 2\alpha)\eta}{a_{red.} - 1 + \epsilon a\left[1 - a_{red.}\eta + \alpha\eta(a_{red.} - 1)\right]} \tag{42}$$

where the abbreviation $\eta = 1 - [1 + a_{red.}]\langle|z|\rangle^{(-)}(1)$ has been introduced. The right hand side depends on $\alpha$ in a monotonous way (cf. appendix A). Therefore for $\alpha$ values in a neighbourhood of $\alpha = 1/2$ eq.(42) determines the unique $c$ value in the allowed interval $[1/a_{red.}, a_{red.}]$ which satisfies eq.(27) by construction. Hence every density of the form (33) yields a period two orbit of the mean field equation (20). The fixed point solution which also exists at the parameter values under consideration is contained in this continuous family for a particular value of the parameter $\alpha$

*Higher order periodic orbits*: It is obvious that the computations made above can be applied to any band merging in the tent map (e.g. consider the ergodic component $\rho^{(-)}$). Therefore at the critical parameter values $(1-\epsilon)a = 2^{1/2^n}$ a continuum of period $2^n$ orbits occurs. Every solution is stable. Depending on the initial condition a particular state is attained in the course of the dynamics.

## 4  Analysis of coupled logistic equations

The most popular system which is discussed in the literature is given by coupled logistic equations
$$f(x) = 1 - ax^2 \quad . \tag{43}$$
Although the dynamics of the uncoupled map is fairly well understood on a rigorous level [29] it is surprisingly difficult to obtain related results for the mean field coupled map lattice. The main problem originates from the non hyperbolicity of the map which causes the dense set of windows of stable periodic orbits in the uncoupled case. Because of the lack of suitable analytical approaches one is forced to resort to numerical solutions of eq.(5). Nevertheless some statements on stationary states can be made.

*Stationary states*: The mean field map which governs the dynamics of eq.(5) reads
$$T_n(x) = 1 - \epsilon a \langle x^2 \rangle_n - (1-\epsilon)ax^2 \quad . \tag{44}$$

To analyze the stationary states it is again useful to introduce rescaled phase space variables so that eq.(44) takes the form of the logistic equation. Choosing
$$\gamma_{n+1} = 1 - \epsilon a \gamma_n^2 \langle z^2 \rangle_n \tag{45}$$
the mean field equation is cast into the form (19), (20) with
$$\tilde{T}(z) = 1 - \frac{a_{red.}\gamma_n^2}{1 - \epsilon a \gamma_n^2 \langle z^2 \rangle_n} z^2 \tag{46}$$

where the brackets $\langle \ldots \rangle_n$ denote the average with respect to the density (19). Let me focus on the discussion of fixed points. Periodic orbits can be treated in principle by considering the iterated system. However the discussion of the fixed point problem seems to contain the main structure of the whole bifurcation diagram. The fixed point density $\tilde{\rho}_*$ is determined by an equation analogous to eq.(8) where

$$\tilde{T}_*(z) = 1 - a_{red.}\gamma_* z^2 \qquad (47)$$

denotes the mean field map and the scaling parameter obeys

$$\gamma_* = 1 - \epsilon a \gamma_*^2 \langle z^2 \rangle_* \quad . \qquad (48)$$

If one regards for the moment the quantity $c := a_{red.}\gamma_*$ as a free parameter then $\tilde{\rho}_*$ is the unique SRB measure of an ordinary logistic equation[4]. This density fixes the quantity $\langle z^2 \rangle_* = \langle z^2 \rangle_*(c)$. If one now eliminates the scaling parameter $\gamma_*$ from eq.(48) in favour of the abbreviation $c$ one ends up with

$$(1-\epsilon)^2 a = c + \epsilon c \left( c \langle z^2 \rangle_*(c) - 1 \right), \quad c \in [0, 2] \quad . \qquad (49)$$

Eq.(49) yields a family of curves in the parameter plane $(a, \epsilon)$. On every curve the coupled map lattice admits a stationary solution. If the parameter $c$ is chosen in such a way that its value corresponds to a periodic window in the logistic equation then the density $\tilde{\rho}_*$ consists of a finite collection of $\delta$ peaks. Due to to the discussion at the end of section 2 these states are dynamically stable. In the opposite case less can be said about the density $\tilde{\rho}_*$ and their stability. However the remarks made in section 2 indicate that the solutions are dynamically unstable at least for situations which correspond to Smale complete logistic equations.

To sum it up it can be said that eq.(49) provides a foliation which translates the bifurcation structure of the logistic map to the parameter plane $(a, \epsilon)$. States which correspond to periodic windows lead to dynamical stable fixed points. The fixed point analysis presented here has much in common with the approach used in ref.[18]. But in contrast to this approach eq.(49) provides an explicit expression for the bifurcation lines and clarifies the stability of the solutions. Finally it should be mentioned that to my best knowledge it is not clear whether the parametrization (49) depends continuously on $c$[5]. Additionally it is not clear whether the lines can intersect so that coexisting stationary states occur.

*Numerical simulations*: The discussion in the in the preceding paragraph has shown that the bifurcation diagram becomes tremendous complicated even if only the fixed points are considered. Although there exist stable solutions it is not clear whether a typical initial condition settles on these solutions. But it is expected that an infinitesimal coupling changes the dynamical behaviour in contrast to the hyperbolic example treated in the preceding sections.

---

[4]Although every physicist believes that the logistic equation possesses an SRB measure I am not aware of any proof that eq.(43) possesses for *every* $a \in [0, 2]$ a unique measure that describes the evolution of Lebesgue almost all initial points $x$.

[5]Whether $\langle z^2 \rangle_*(c)$ depends continuously on $c$ seems to be an unsolved problem. In fact both hypothesis can be found in the literature[18, 22].

To shed some light on these problems the evolution equation (5) has been iterated numerically. The numerical algorithm which has been used is briefly described in appendix B. Main interest has been focused on the regime of weak and moderate coupling strength. The uniform distribution was chosen as a typical initial density. It corresponds to a random uniform distribution of phase space coordinates for the original coupled map lattice (1). The results described below seem not to depend on this choice. If the parameter $a$ is restricted to a large periodic window one finds in the weak coupling regime stable periodic solutions of the same period. They correspond to the solutions which have been predicted by the analytical approach presented above. More interesting things happen if the parameter is chosen as a "typical" chaotic value. Several results for the choice $a = 2$ are presented here but qualitative similar results are obtained at all other values that have been investigated. Even for small coupling strength the system does not settle to a stationary state. The distribution of the mean field develops a complicated shape which shows an apparent nonsystematic variation with the coupling strength (cf. Fig.2). The width of the distribution is finite and is responsible for the so called "violation of the law of large numbers" [22, 7]. The time evolution of the mean field is conveniently analyzed in terms of its power spectrum (cf. Fig.3). For weak coupling strength the spectrum shows a small noise level on which sharp peaks are superimposed. By increasing the coupling strength the peaks broaden and the noise level increases. The actual shape of the spectrum seems to depend sensitively on the coupling strength. Because of the complicated structure of the partial bifurcation diagram described above it seems to be hopeless to attribute the peaks to definite periodic orbits. Finally Fig.4 contains the time evolution of the density $\rho_n$ for two values of the coupling strength. In the case of weak coupling the density looks like a stochastic perturbation of the invariant density of the uncoupled system. The case of moderate coupling for which a broad band noise in the power spectrum of the mean field is observed yields a density with a strongly developed structure. It possesses sharp peaks which show an intermittent time evolution. The appearance of localized peaks in the density indicate a clustered state in the original coupled map lattice. Hence this type of motion can be viewed as an intermittent dynamics of clusters in the original system. Qualitatively this type of motion can be attributed to the stable period three window in the logistic equation. Indeed for slightly different coupling strength a stable period three state can be found in numerical simulations of the mean field equation.

<Fig.2

<Fig.3

<Fig.4

# 5  Conclusion

The simplicity of the global coupling has allowed for a mean field like description of the coupled map lattice. It contains the limit of infinite system size in a quite simple way namely by a "smoothing" process of the reduced density. In a certain sense this description circumvents the problem of supertransients. If the transient time increases tremendous with the system size, e.g. exponentially, then the relevant dynamics of the spatially extended system may be the transient one and not the (mathematical) stationary state. This behaviour is of course difficult to analyze analytically for the original system (1). One merit of the mean field formulation consists in the fact that in the limit of infinite system size, that means among continuous densities, these transients correspond to stationary states which can be discussed more easily.

For the example of tent maps a fairly complete survey over the stationary states has been given. It is worth to mention that these solutions can be found in numerical solutions of the mean field equation as well as direct simulations of the coupled map lattice. They are attained for all initial conditions so that these states are apparently not only locally but also globally stable. Hence the stationary state is a unique fixed point for $2 \geq (1-\epsilon)a > \sqrt{2}$. The finite size effects induce fluctuations around this fixed point so that global quantities like the mean field "obey the law of large numbers". At other parameter values one solution is selected from the continuum of periodic states depending on the chosen initial condition. Because of its periodicity the mean square deviation of global quantities saturates at a finite value and apparently "violates the law of large numbers". The value of saturation depends on the initial condition (cf. [16]). At different parameter values that are not covered by the present approach, especially at negative coupling, additional bifurcations may occur [22]. These values have been skipped in the present discussion because the map lattice may have diverging trajectories.

Beyond the phenomena encountered in the hyperbolic tent map the non hyperbolic logistic equation shows additional features. First of all the degeneracy of the spectrum of the transfer operator causes, that even an infinitesimal coupling changes the dynamics drastically. This is obvious from the bifurcation analysis of fixed points. In contrast to the hyperbolic situations the numerical simulations indicate that the locally stable solutions are not attained for a typical initial condition. Hence their domain of attraction seems to be small. In view of the complicated structure of the bifurcation diagram with probably infinitely many coexisting stable states this observation is not very astonishing. The lack of a unique stable fixed point also explains that the phenomenon of the "violation of the law of large numbers" is typically observed in non hyperbolic examples at small coupling strength.

Finally it should be mentioned that it is difficult to decide which features survive if the condition of infinite coupling range is relaxed. It is however remarkable that hyperbolic coupled maps show often a rather simple evolution whereas non hyperbolic maps show a complicated dynamics with many coexisting stable or metastable states and intermittent behaviour. The investigation of globally coupled systems may shed additional light on these phenomena. But further investigations are required.

## Acknowledgement

The author is indebted to the Deutsche Forschungsgemeinschaft for financial support by a "Habilitandenstipendium". This work was performed within a program of the Sonderforschungsbereich 185 Darmstadt–Frankfurt, FRG.

## Appendix A

*Proof of eq.(32)*: The abbreviation

$$g(z; 1/c) := 1 - \frac{a_{red.}}{c}|z| \qquad (50)$$

is introduced to take explicitly the $c$ dependence into account. Obviously the relations

$$\tilde{T}_1^*(z) = g(z; 1/c), \quad \tilde{T}_2^*(z) = g(z; c) \qquad (51)$$

hold. The ergodic components $\rho^{(\pm)}(z; c)$ are the continuous invariant measures of the map $g(g(z; 1/c); c)$. Hence the functions

$$\psi^{(\pm)}(z) := \int \delta\left(z - g(z'; 1/c)\right) \rho^{(\pm)}(z'; c) \, dz' \qquad (52)$$

are continuous invariant densities of the map $g(g(z; c); 1/c)$ which coincides with the former one if $c$ is replaced by $1/c$. This map has two ergodic components which are contained in the two intervals $[-\zeta(1/c), \zeta(1/c)]$ respectively $[\zeta(1/c), 1]$. The original densities $\rho^{(\pm)}$ are contained in the intervals $[-\zeta(c), \zeta(c)]$ respectively $[\zeta(c), 1]$. One easily calculates that the images of these intervals obey

$$g([-\zeta(c), \zeta(c)]; 1/c) = [\zeta(1/c), 1], \quad g([\zeta(c), 1]; 1/c) \subseteq [-\zeta(1/c), \zeta(1/c)] \qquad (53)$$

Hence the densities (52) are continuous functions on the intervals (53) and coincide for that reason with the ergodic components $\rho^{(\mp)}(z; 1/c)$.

*Relation between* $\langle|z|\rangle^{(-)}$ *and* $\langle|z|\rangle^{(+)}$: The case $c = 1$ will be considered throughout this paragraph and the argument $c = 1$ will be suppressed in the notation. On the two different ergodic components the map (28) which is simply the second iterate of the tent map reads

$$g^{(-)}(z) := 1 - a_{red.}(1 - a_{red.}|z|), \quad |z| < \zeta \qquad (54)$$
$$g^{(+)}(z) := 1 - a_{red.}|1 - a_{red.}z|, \quad \zeta < z \leq 1 \quad . \qquad (55)$$

Both maps are conjugate to each other via $g^{(+)} \circ h = h \circ g^{(-)}$ where $h(z) = (1-z)/a_{red.}$. On one hand the conjugacy implies

$$\int z \rho^{(-)}(z)\, dz = \int h^{-1}(z) \rho^{(+)}(z)\, dz = 1 - a_{red.} \int z \rho^{(+)}(z)\, dz \quad . \tag{56}$$

On the other hand the invariance of the density $\rho^{(-)}$ yields

$$\int z \rho^{(-)}(z)\, dz = \int g^{(-)}(z) \rho^{(-)}(z)\, dz = 1 - a_{red.} + a_{red.}^2 \langle |z| \rangle^{(-)} \quad . \tag{57}$$

If one combines both equations one obtains

$$\langle |z| \rangle^{(+)} = 1 - a_{red.} \langle |z| \rangle^{(-)} \quad . \tag{58}$$

*Discussion of eq.(42)*: It will be shown that eq.(42) determines a unique non negative value for every admissible $\alpha$ value. Because $\rho^{(-)}(z;1)$ is contained in the interval $[-\zeta(1), \leq \zeta(1)]$ the inequality $\langle |z| \rangle^{(-)}(1) \leq \zeta(1) = 1/(a_{red.}+1)$ holds. But then $0 \leq \eta \leq 1$ is valid. The denominator of eq.(42) obeys for $\alpha \in [0,1]$

$$a_{red.} - 1 + \epsilon a(1 - a_{red.}) \leq a_{red.} - 1 + \epsilon a[1 - a_{red.}\eta + \alpha\eta(a_{red.} - 1)] \quad . \tag{59}$$

Hence it is positive for $\epsilon < 1/2$ (this bound on the coupling strength can be relaxed if a better estimate of the quantity $\eta$ would be derived). But then the right hand side of eq.(42) is a monotonous function so that for every $\alpha \in [0,1]$ a unique value of $c$ is found. This value is positive because

$$\begin{aligned} c &\geq 1 - \frac{\epsilon a(a_{red.} - 1)\eta}{a_{red.} - 1 + \epsilon a[1 - a_{red.}\eta + \eta(a_{red.} - 1)]} \\ &= \frac{(a_{red.} - 1)(1 - \epsilon a)}{a_{red.} - 1 + \epsilon a[1 - a_{red.}\eta + \eta(a_{red.} - 1)]} > 0 \end{aligned} \tag{60}$$

holds for $\epsilon < 1/2$. From the estimate

$$\left| \frac{\epsilon a(a_{red.} - 1)(1 - 2\alpha)\eta}{a_{red.} - 1 + \epsilon a[1 - a_{red.}\eta + \alpha\eta(a_{red.} - 1)]} \right| \leq \frac{\epsilon a}{1 - \epsilon a} \tag{61}$$

one additionally obtains that for sufficiently small coupling strength $\epsilon$ any value $\alpha \in [0,1]$ corresponds to a unique value $c \in [1/a_{red.}, a_{red.}]$.

## Appendix B

Let me briefly describe the basic idea for the numerical simulation of eq.(5). The straightforward approach consists in an equidistant partition of the interval and the approximation of the density by a piecewise constant function in every iteration step. But an equidistant partition seems to be not very suitable especially if the density develops square root like singularities. A finer partition in those regions is desirable whereas in flat regions a coarser partition is sufficient. For that reason the densities are approximated by piecewise constant functions so that every interval contains the same weight. To be definite suppose that at time $n$ a partition $I_k^{(n)} = [a_{k-1}^{(n)}, a_k^{(n)}]$, $1 \leq k \leq N$ in $N$ intervals is given so that the density has the form

$$\rho_n(z) = \sum_{k=1}^{N} \frac{1}{N \Delta a_k^{(n)}} \chi_{I_k^{(n)}}(z) \tag{62}$$

where $\Delta a_k^{(n)} := a_k^{(n)} - a_{k-1}^{(n)}$ denote the lengths of the intervals and $\chi_J$ stands for the characteristic function of the interval $J$. The partition at time $n+1$ is determined in such a way that the density $\rho_{n+1}$ yields the same weight $1/N$ for every interval. This prescription results in

$$\begin{aligned}
\frac{1}{N} &= \int_{a_{k-1}^{(n+1)}}^{a_k^{(n+1)}} \int \delta(z - T_n(z')) \rho_n(z') \, dz' \, dz \\
&= \sum_{l=1}^{N} \int_{a_{k-1}^{(n+1)}}^{a_k^{(n+1)}} \int_{a_{l-1}^{(n)}}^{a_l^{(n)}} \delta(z - T_n(z')) \, dz' \, dz \frac{1}{N \Delta a_l^{(n)}}
\end{aligned} \tag{63}$$

The partition at time $n+1$, $\{a_k^{(n+1)}\}$, can be determined easily from eq.(63) in an iterative way because the remaining integrals are nothing else but the lengths of the intervals $I_l^{(n)} \cap T_n^{-1}(I_k^{(n+1)})$[6]. At time $n+1$ the density is again approximated by a piecewise constant function on the new partition (cf. eq.(62)) which corresponds formally to the canonical orthogonal projection.

Numerical estimates show that the error of the algorithm is of the order $O(1/N)$ even if the density develops singularities. For the numerical simulations a partition of size $N = 2000$ has turned out to be sufficient so that the computations can be performed on every PC.

---

[6]An explicit algorithm for symmetric unimodal maps is available on request.

# Figure captions

Fig.1 Diagrammatic view of the two times iterated mean field map (28) for $\sqrt[4]{2} < a_{red.} < \sqrt{2}$. The diagonal, the dashed boxes, and the unstable fixed point $\zeta(c)$ are indicated for clearness.

Fig.2 Distribution function of $\langle x^2 \rangle_n = (1 - h_n)/a$ for coupled logistic equations at $a = 2$ and several coupling strengths. The distributions have been obtained from a time series of $10^5$ data points. They are displayed as histograms with 200 boxes on the abscissa. The size of the boxes are adapted to the width of the distribution.

Fig.3 Power spectra of $\langle x^2 \rangle_n$ for coupled logistic equations at $a = 2$ and several coupling strengths .The spectra have been obtained from the Fourier transform of a time series of length 1024. An average over 200 different series has been performed.

Fig.4 Time evolution of the reduced density $\rho_n(x)$ for coupled logistic equations at $a = 2$ and two values of the coupling strength (a) $\epsilon = 0.01$, (b) $\epsilon = 0.2$. The densities are displayed as histograms with 100 boxes on the $x$ axis.

Fig. 1

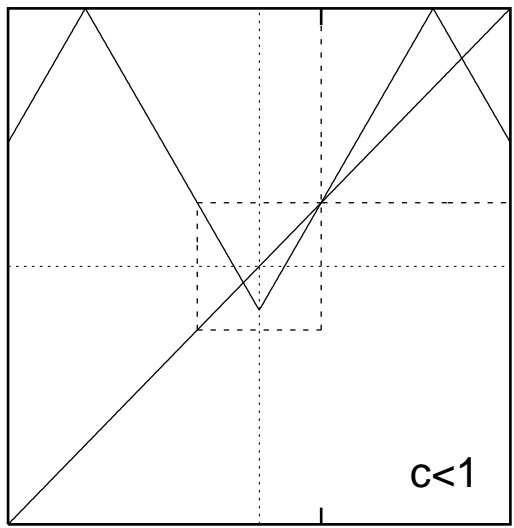 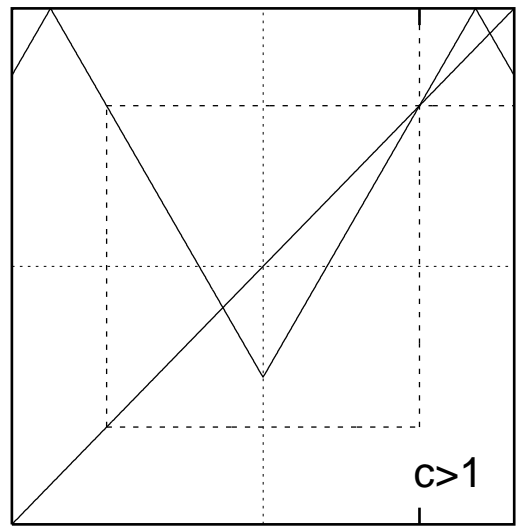

ζ(c)     ζ(c)

Fig. 2

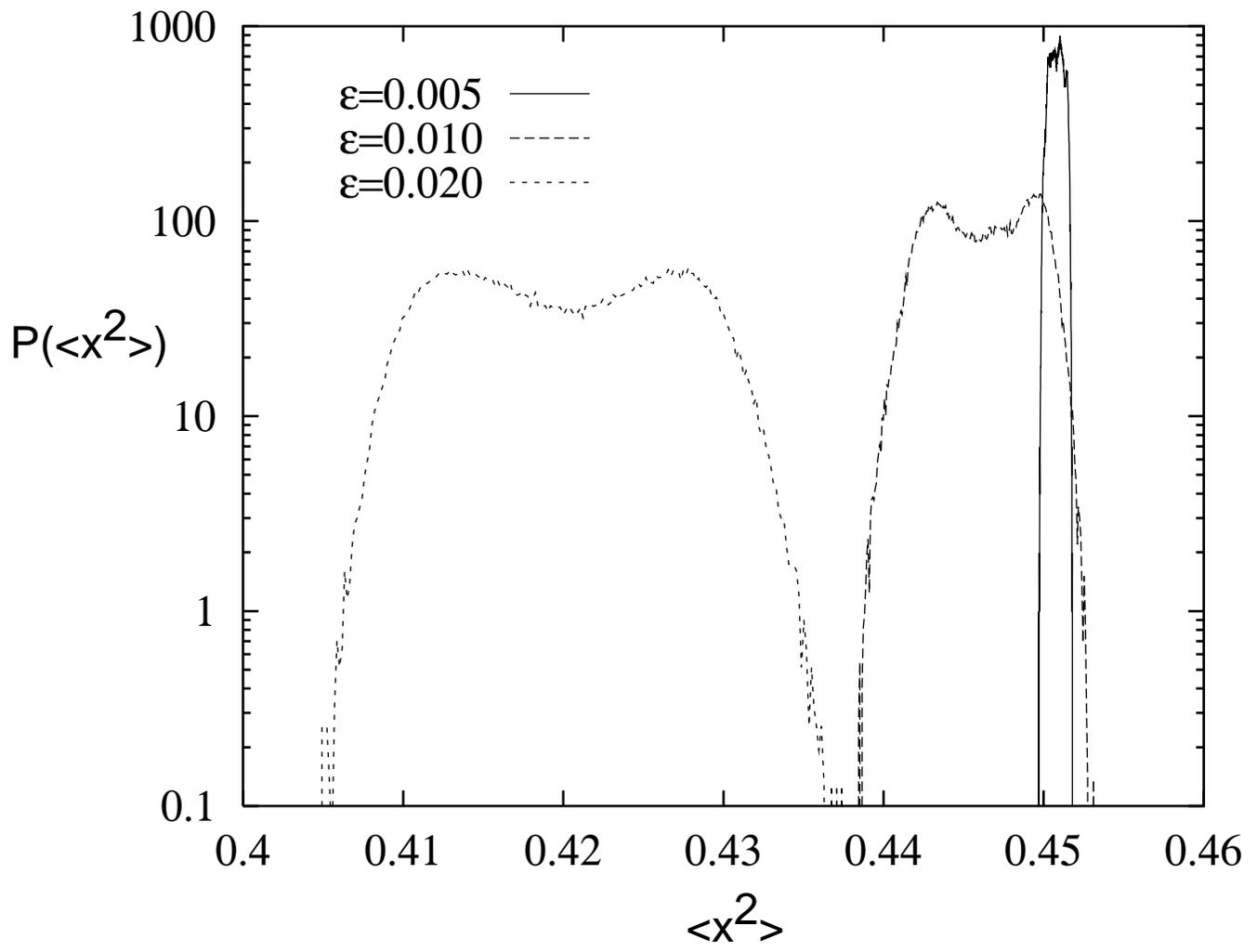

Fig. 3

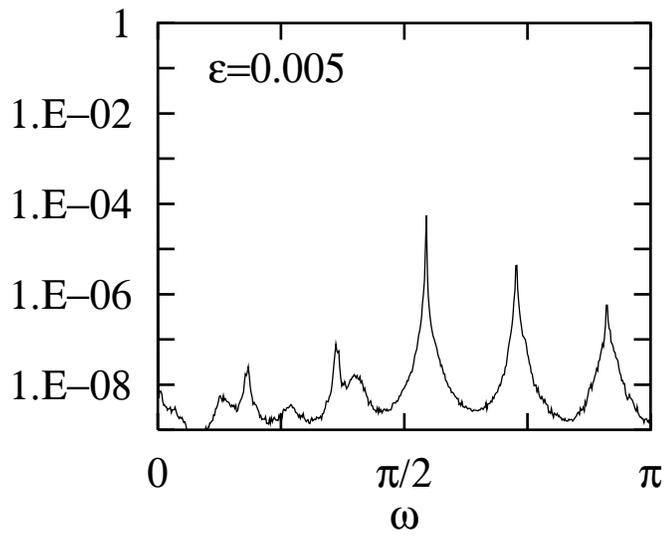
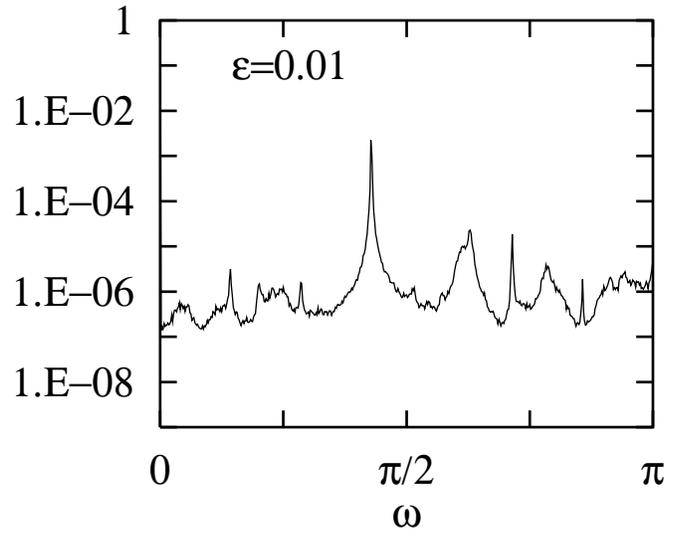
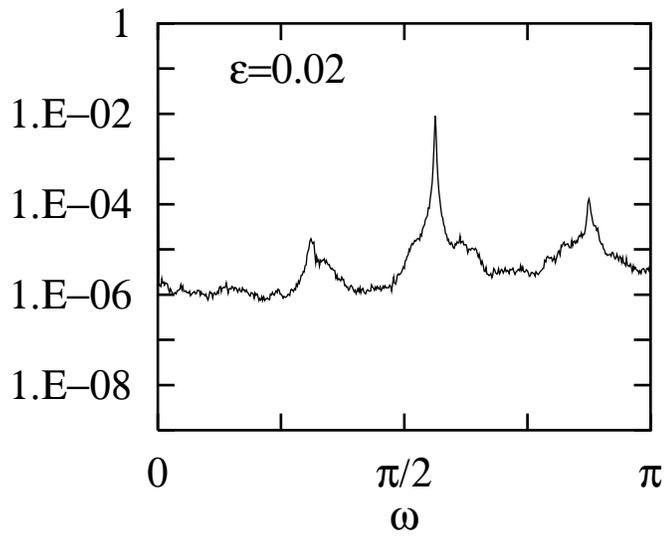
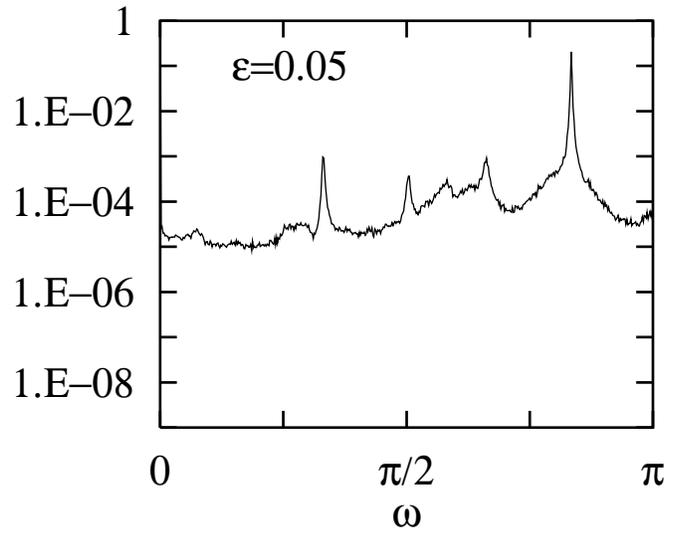
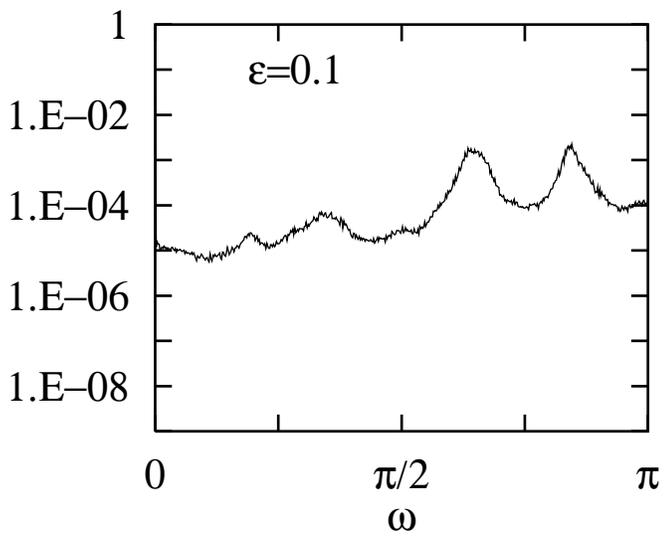
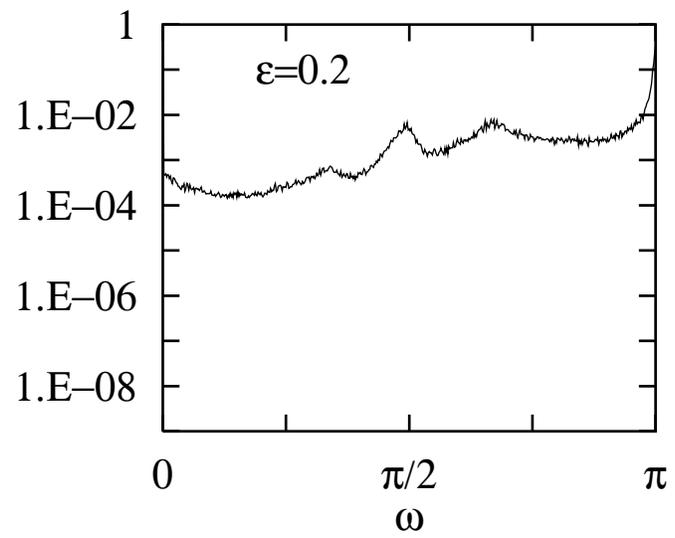

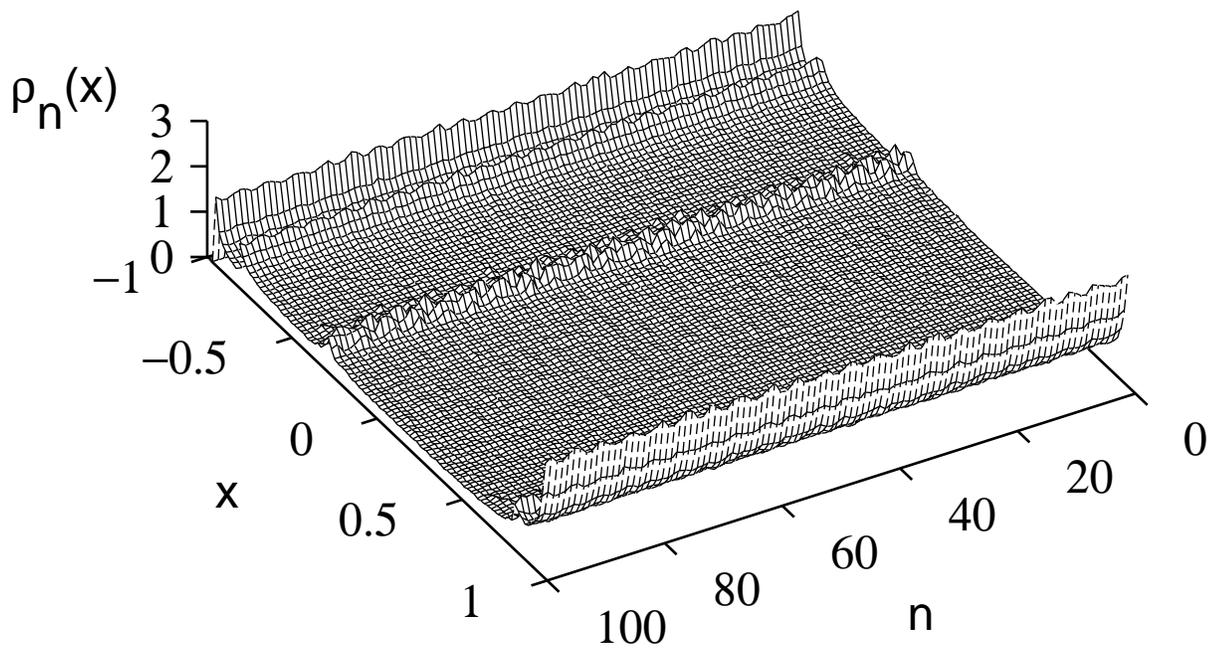

Fig. 4a

Fig. 4b

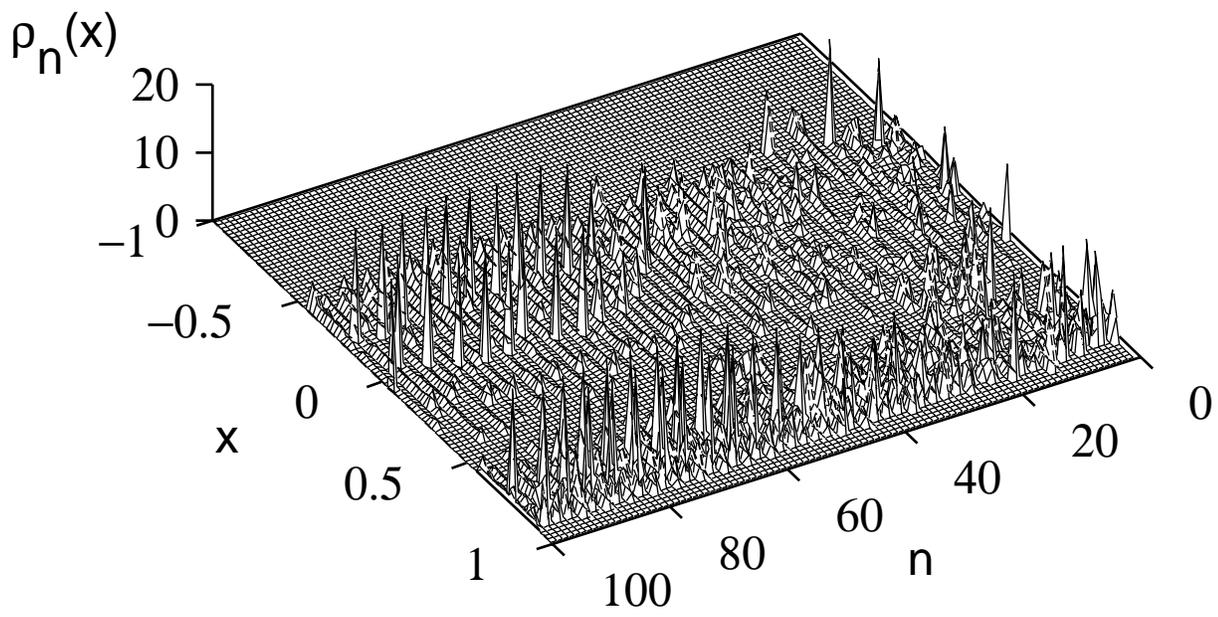